\documentclass[aps,twocolumn]{revtex4}
\usepackage{graphics}

\begin{document}
\title{Two-photon absorption in potassium niobate}

\author{A. D. Ludlow, H. M. Nelson, and S. D. Bergeson \email{scott.bergeson@byu.edu}
}

\affiliation{Department of Physics and Astronomy, Brigham Young
University, Provo, UT  84602}

\begin{abstract}
We report measurements of thermal self-locking of a Fabry-Perot
cavity containing a potassium niobate (KNbO$_3$) crystal.  We
develop a method to determine linear and nonlinear optical
absorption coefficients in intracavity crystals by detailed
analysis of the transmission lineshapes.  These lineshapes are
typical of optical bistability in thermally loaded cavities. For
our crystal, we determine the one-photon absorption coefficient at
846 nm to be $\alpha = (0.0034 \pm 0.0022)~\mbox{m}^{-1}$ and the
two-photon absorption coefficient at 846 nm to be $\beta = (3.2
\pm 0.5) \times 10^{-11} ~\mbox{m/W}$ and the one-photon
absorption coefficient at 423 nm to be $(13 \pm 2)$ m$^{-1}$. We
also address the issue of blue-light-induced-infrared-absorption
(BLIIRA), and determine a coefficient for this excited state
absorption process.  Our method is particularly well suited to
bulk absorption measurements where absorption is small compared to
scattering.  We also report new measurements of the temperature
dependence of the index of refraction at 846 nm, and compare to
values in the literature.
\end{abstract}


\maketitle

\section{Introduction}

For cw laser second harmonic generation in a build-up cavity,
thermal issues in the nonlinear crystal inevitably come into play.
This is particularly true at high laser powers.  In typical
applications, a Fabry-Perot build-up cavity increases the
fundamental laser intensity by one or two orders of magnitude
compared to the single-pass intensity.  The nonlinear crystal
absorbs a fraction of the fundamental power, heats up, and changes
the cavity properties.  Thermal lensing changes the cavity's
spatial mode.  Nonlinear absorption changes the cavity's finesse.
Thermal expansion and changes in the crystal's index of refraction
change the cavity's optical path length.

These effects are well known, and have been observed by several
groups working in this field (for example,
\cite{Arie92,Polzik91}). Thermally-induced changes in the build-up
cavity sometimes limit the second-harmonic generation efficiency.
In many cases, they are nuisances that can be worked around.
However, these changes can be used as a sensitive diagnostic tool
for accurately characterizing nonlinear crystal properties.

For gases and atomic vapors, measuring optical properties of
materials inside Fabry-Perot cavities has a long history
\cite{Siegman86}, particularly in relation to optical bistability,
saturable absorption, and index of refraction measurements.
Thermal and collisional properties of gases have also been
measured in this way \cite{Ye96}, as well as optical properties of
cavity mirrors \cite{An97}. However, measurements of optical
properties of crystals in Fabry-Perot cavities is relatively new
\cite{Douillet99}.

In some ways, Fabry-Perot measure\-ments are similar to la\-ser
cal\-o\-rim\-e\-try \cite{Busse94}.  In those cal\-orimetry
measurements, a laser passes through a thermally isolated
absorbing crystal and heats it slightly.  From the temperature
rise, laser power, thermal mass, and thermal conductivity it is
possible to determine an optical absorption coefficient for the
crystal.  Comparing calorimetry to our Fabry-Perot measurements,
both are photothermal methods.  Both can have high sensitivities
to small absorptions.    In contrast to calorimetry, our
``thermometer'' is the power-induced shift in the cavity resonance
frequency.  Fabry-Perot measurements can typically reach much
higher intensities, making nonlinear effects more apparent.
Furthermore, frequency-based techniques, such as the one presented
in this paper, usually offer a higher level of precision.

This paper presents a detailed study of the thermal response of
KNbO$_3$ inside a Fabry-Perot build-up cavity.  We use this
response to determine a new value for the optical absorption
coefficients in KNbO$_3$ at 846 nm and 423 nm, and to measure the
effects of blue-light-induced infrared absorption (BLIIRA).  We
also measure the temperature-dependence of the index of refraction
for light polarized along the $b$ and $c$ crystal axes.  These
measurements are compared to values from the literature.

\section{Experimental Setup}
\begin{figure}[t]
\centerline{\rotatebox{270}{\scalebox{.31}{\includegraphics{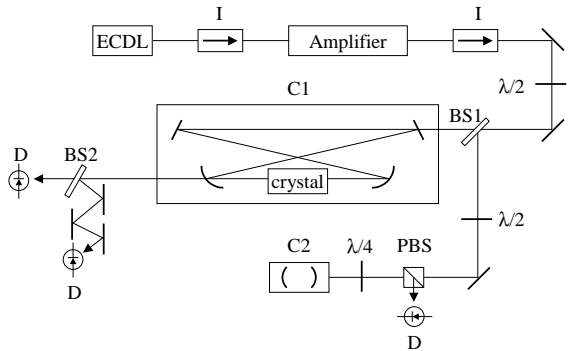}}}}
\caption{Schematic diagram of the experimental layout.  ECDL =
extended cavity diode laser, I = optical isolator, $\lambda/2$ =
half wave plate, $\lambda/4$ = quarter wave plate, C1 = frequency
doubling ``bowtie'' cavity, C2 = stable optical cavity, D =
photodiode detectors, PBS = polarizing beam-splitter cube, BS1 =
uncoated quartz window, BS2 = dichroic beam splitter.  The short
heavy lines are mirrors.
} \label{fig:expLayout}
\end{figure}

Our experimental setup is typical for cw second-harmonic
generation in an external build-up cavity (see Figure
\ref{fig:expLayout}).  This laser system is designed for a calcium
laser cooling experiment at 423 nm. It produces typically 75 mW at
423 nm, measured outside the cavity.

The laser system consists of a single frequency extended cavity
diode laser at 846 nm amplified in a single-pass through a tapered
laser diode \cite{NewFocus}.  The extended cavity diode laser is
frequency-stabilized relative to a separate optical cavity using
the Pound-Drever method \cite{Pound83} to provide both short and
long-term stability.  After optical isolation, polarization
correction, and mode matching, we inject 200 mW into the Gaussian
mode of a four-mirror folded (bowtie) Fabry-Perot frequency
doubling cavity.  Higher-order modes in the cavity are less than
5\% of the Gaussian mode.  The input coupler reflectivity is
96.6\%, and the other three mirror reflectivities are $>99.5$\% in
the infrared.  The radius of curvature for the two curved mirrors
is 100 mm, and they are separated by a distance of 117 mm.  The
angle of incidence on the curved mirrors is
$6.5^{\mbox{\footnotesize o}}$, minimizing the ellipticity of the
beam waist inside the crystal (41.1 and 40.7 $\mu$m in the
tangential and saggital planes, as determined from the cavity
geometry).  The round-trip cavity length is 618 mm. The KNbO$_3$
crystal \cite{VLOC} is $a$-cut and antireflection-coated
($R<$0.25\%) at 846 nm.  The non-critical phase-matching
temperature is $-11.5^{\mbox{\footnotesize o}}$C, for light
polarized along the $b$-axis.  The crystal is mounted on a
Peltier-cooled copper block.  An AD590 temperature sensor and a
thermister embedded in the copper mount monitor the crystal
temperature.  The entire bowtie cavity is enclosed in a sealed
aluminum box, filled with dry oxygen at atmospheric pressure.  For
low laser powers, and when the crystal temperature is far from the
phase matching temperature, the cavity finesse with the crystal in
place is 124. Without the crystal, the cavity finesse is 140.

\section{Transmission Lineshape Without Blue Light}

When laser light incident on the bowtie cavity is polarized along
the $c$-axis, it is impossible to meet the phase-matching
condition for non-critical second harmonic generation at
$\lambda=846$ nm, and no blue light is generated.  Previous
measurements in KNbO$_3$ have shown that even small amounts of
blue light dramatically change the infrared absorption properties
\cite{Goldberg93,Mabuchi94,Shiv95}. We address this
blue-light-induced-infrared-absorption (BLIIRA) later in the
paper.  However, in this section we discuss infrared absorption in
the {\it absence} of blue light, {\it i.e.}: when the incident
laser beam is polarized along the $c$-axis.

One of the flat mirrors in the bowtie cavity is mounted on a
piezoelectric crystal, and we use this crystal to change the
optical length of the cavity.  A small amount of infrared light is
transmitted through the cavity mirrors, and we measure this light
while sweeping the cavity length.  When the piezo-mounted mirror
passes through the position for cavity resonance ({\it i.e.}: when
the roundtrip pathlength in the cavity is an integer multiple of
the laser wavelength), the power circulating in the cavity, and
transmitted through the cavity end mirror, increases.  The
transmission lineshape is Lorenzian in the absence of disturbing
effects:

\begin{equation}
P(\nu)={{P_0}\over {1+{F}\left(\nu-\nu_0 \right)^2}},
\label{eqn:lorentz}
\end{equation}

\noindent where $\nu_0$ is the
resonant frequency of the cavity, $2 / \sqrt{F}$ is the full
width at half-maximum (FWHM), and $P_0$ is the infra-red power
at the peak of the Lorenzian.

The KNbO$_3$ crystal inside the bowtie cavity absorbs a small
amount of the laser power and heats up.  This changes the
crystal's length and index of refraction, which in turn changes
the optical path length of the bowtie cavity.  We model the
overall effect by replacing $\nu_0$ with $\nu_0 + \eta P/P_0.$
\noindent Accordingly, Equation \ref{eqn:lorentz} becomes

\begin{equation}
{y = {{1} \over {1 + F \left( \nu - \eta y \right)^2}}}.
\label{eqn:final_trans}
\end{equation}

\noindent where $y = P/P_0$ is the normalized transmission, and
the frequency $\nu$ is measured relative to the unshifted
frequency $\nu_0$.  Transmission lineshapes of this kind are
typical of optically bistable devices.  A similar treatment of
this effect is given in the references \cite{Douillet99}.

\begin{figure}[t]
\centerline{\rotatebox{90}{\scalebox{.31}{\includegraphics{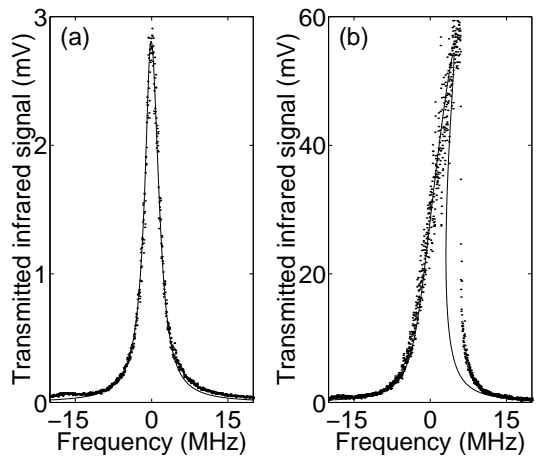}}}}
\caption{Infrared transmission lineshape for the bowtie cavity in
the absence of blue light.  The points are the measurements.  The
solid line is a fit to the data (see Equation
\ref{eqn:final_trans}).  The infrared light is polarized along the
$c$-axis.  (a) Low power.  (b) High power.  For high powers, the
fit function is not single-valued above the cavity resonant
frequency.
} \label{fig:lineshape}
\end{figure}

The peak in the transmission curve occurs when $y=1$ and
$\nu=\eta$.  Figure \ref{fig:lineshape} shows the transmission
lineshape for both low and high laser powers.  On the same figure
we plot a fit of the data to Equation \ref{eqn:final_trans}.  The
asymmetry in the transmission lineshape at high powers is due to
thermal changes in the optical pathlength.  When sweeping the
cavity from low to high frequencies as in Figure
\ref{fig:lineshape}, the power-dependent cavity resonance is
effectively ``dragged along'' from $\nu = \nu_0$ to $\nu = nu_0 +
\eta$ as the cavity length changes.  This is known as ``thermal
self-locking'' \cite{Ye96}.  The value of $\eta$ is called the
``self-locking range,'' and it is related to the ``thermo-optic
constant'' of reference \cite{Douillet99}.

For a given power incident on the cavity, we determine the
self-locking range $\eta$ using a nonlinear least-squares fit of
the measured transmission lineshape to Equation
\ref{eqn:final_trans}.  For the highest laser intensities, it
appears that a better overall fit is obtained by replacing $\eta
y$ in Equation \ref{eqn:final_trans} by a higher order polynomial.
However, the fit in these high intensity data is hampered because
the fit function is not single-valued, as can be seen in Figure
\ref{eqn:final_trans}.  For these measurements, we determine  the
values of $\eta$ visually by finding the frequency at the peak of
the transmission curves.

\begin{figure}[t]
\centerline{\rotatebox{90}{\scalebox{.31}{\includegraphics{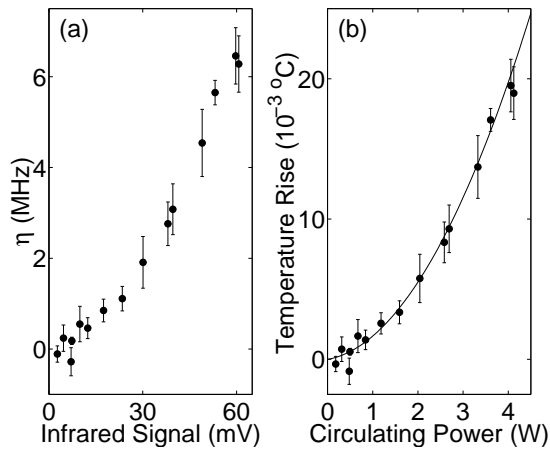}}}}
\caption{(a) Self-locking range, $\eta$, versus infrared detector
signal.  (b) The data from \ref{fig:slr}a with temperature and
power calibrations.  The solid line is a weighted fit to the data,
$T = c_1 P + c_2 P^2$.
} \label{fig:slr}
\end{figure}

Our values for $\eta$ as a function of infrared power in the
bowtie cavity are shown in Figure \ref{fig:slr}a.  The error bars
indicate the 1 $\sigma$ statistical spread in 5 measurements at
each power setting.  For linear absorption, the cavity shift,
$\eta$, should be a linear function of the circulating power.
However our measurements indicate a quadratic dependence of the
cavity shift on the circulating power, suggesting that the
dominant absorption mechanism is two-photon absorption in the
infrared.

We convert $\eta$ into a temperature rise inside the crystal by
measuring the temperature change necessary to shift the cavity
resonance frequency by one free spectral range.  In these
calibration measurements, we attenuate the incident laser power to
10 $\mu$W to avoid laser-heating in the crystal.  The cavity
mirror positions remain fixed, and the laser frequency is
constant.  Using the Peltier device, we change the temperature of
the crystal by several degrees while monitoring the cavity
transmission.  At $T=-11.5^{\mbox{\footnotesize o}}$C, the
resonant frequency of the bowtie cavity shifts at a rate of 231
MHz $^{\mbox{\footnotesize o}}$C$^{-1}$ for light polarized along
the $b$-axis and 331 MHz $^{\mbox{\footnotesize o}}$C$^{-1}$ for
light polarized along the $c$-axis.

We calibrate our photodiode signal using a power meter, and
measure the efficiency of the optical path to convert this IR
signal to a power circulating inside the cavity. The conversion
from cavity shift to temperature rise can also be derived from
measurements of the change in the index of refraction with
temperature.  From the published values of $dn/dT$ \cite{Ghosh94}
and the geometry of our bowtie cavity, we calculated what the
temperature-dependent shift ought to be.  The calculation agrees
with our data for the $b$-axis.  However, we find a small
difference between the calculation and our data for the $c$-axis,
as discussed later in this paper.

Using our measurements, we convert $\eta$ into $\Delta T$, the
difference in temperature between the crystal axis where the laser
propagates and the crystal wall which is kept constant by our
temperature control circuit.  Figure \ref{fig:slr}b shows a plot
of the temperature rise as a function of the infrared power
circulating inside the cavity.  It is relatively simple to convert
this temperature rise into a measure of the optical absorption
coefficient of KNbO$_3$ by solving the heat transfer equation,
which we do below.

\section{Solving the Heat Transfer Equation}

We solve the heat transfer equation in cylindrical coordinates.
The laser beam is in the TEM$_{00}$ Gaussian mode, with a 41
$\mu$m waist.  The confocal parameter of the beam is 6.2 mm, and
at the face of the crystal, the beam waist is only 7\% larger than
in the center of the crystal.  So we approximate the laser beam as
axially symmetric and Gaussian with a 41 $\mu$m waist along the
entire 10 mm length of the crystal.  Our crystal has a square
cross section, 3 mm on a side.  However, because it is so much
larger than the laser beam waist, we assume that the crystal is
cylindrical also, with a 1.5 mm radius.  With these assumptions,
the equilibrium temperature inside the crystal is determined by
solving the heat equation in the form:

\begin{equation}
{1\over r}{d\over dr} \left(r{d u\over d r} \right) = -{{G}\over {K}},
\label{eqn:heat_equation}
\end{equation}

\noindent where $u(r)$ is the temperature as a function of radius,
$G$ is the power density, and $K$ is the thermal conductivity.
Following Sutherland \cite{Sutherland96}, the absorbed power
density inside the crystal can be written as

\begin{equation}
G = \alpha I + \beta I^2
\label{eqn:g}
\end{equation}

\noindent where $I=(2P/\pi a^2)\exp(-2r^2/a^2)$ is the laser
intensity, $P$ is the laser power incident on the crystal, $a$ is
the Gaussian beam waist, $\alpha$ is the linear optical absorption
coefficient, and $\beta$ is the two-photon absorption (TPA)
coefficient.

We can simplify Equation \ref{eqn:heat_equation} by letting
$r=ay/\sqrt{2}$, $A=(P\alpha/2\pi K)$, and $B=(P^2 \beta / 2 \pi^2
a^2 K)$.  With these substitutions, Equation
\ref{eqn:heat_equation} integrates once to

\begin{equation}
{d u\over d y} = {A \over y} \left[\exp(-y^2) - 1 \right]
+ {B \over y} \left[\exp(-2y^2) - 1 \right],
\end{equation}

\noindent where the constant of integration is chosen to avoid a
singularity as $y \rightarrow 0$.  The solution to this
differential equation is

\begin{equation}
u(y) = -{{A}\over{2}} E_1(y^2)  - {{A}\over{2}} \ln(y^2)
       -{{B}\over{2}} E_1(2y^2) - {{B}\over{2}} \ln(y^2) + C
\label{eqn:soln}
\end{equation}

\noindent where $E_n(x)$ is the exponential integral
\cite{Abramowitz64}, and $C$ is the constant of integration.  This
constant is determined by requiring the temperature at the wall of
the crystal to be $T_c$ ({\it i.e.}: $u(r=b)=T_c$), which we
control in the experiment.  With this constraint, we can evaluate
the temperature rise on axis inside the crystal.  The series
expansion for the exponential integral has a logarithmic term that
exactly cancels the $\ln(y^2)$ terms in Equation \ref{eqn:soln}.

\begin{widetext}

\begin{eqnarray}
\Delta T  & = & u(0) - T_c
          =  {{A}\over{2}}
\left[\ln\left({{2b^2} \over{a^2}}\right) + \gamma \right] +
{{B}\over{2}} \left[\ln\left({{2b^2} \over{a^2}}\right) + \gamma +
\ln(2) \right] \nonumber \\
        & = & {{\alpha}\over{4 \pi K}}
        \left[\ln\left({{2b^2}\over{a^2}}\right) + \gamma \right] P
        + {{\beta}\over{4 \pi^2 a^2 K}}
        \left[\ln\left({{4b^2}\over{a^2}}\right) + \gamma \right]
        P^2 , \label{eqn:final_sol}
\end{eqnarray}

\end{widetext}

\noindent where $\gamma=.57721\ldots$ is Euler's constant.

We can analyze the data in Figure \ref{fig:slr} to extract the
linear and two-photon absorption coefficients.  A weighted fit to
the function $T(P) = c_1 P + c_2 P^2$, gives $c_1 = (5.7 \pm 3.6)
\times 10^{-4}~ ^{\mbox{\footnotesize o}}$C W$^{-1}$ and $c_2 =
(1.09 \pm 0.12) \times 10^{-3}~ ^{\mbox{\footnotesize o}}$C
W$^{-2}$.  The uncertainties in these numbers are the statistical
uncertainty from the weighted fit to the data. Taking the thermal
conductivity from \cite{Busse94} $K = 4.0$ W m$^{-1} ~
^{\mbox{\footnotesize o}}$C$^{-1}$, $a = 41 \times 10^{-6}$ m, $b
= 1.5 \times 10^{-3}$ m, and $\gamma = 0.57721$. We determine the
one-photon (linear) absorption coefficient to be $\alpha = (0.0034
\pm 0.0021)$ m$^{-1}$ and the two-photon absorption coefficient to
be $\beta = (3.16 \pm 0.35) \times 10^{-11}$ m W$^{-1}$. The
uncertainties in these absorption coefficients are only
statistical.  We have a systematic uncertainty in the power
measurements of 10\%. Our assumption of cylindrical symmetry in
the crystal for solving the heat equation also introduces an
error.  The error is known, but related to this is the fact that
our laser does not propagate exactly down the center of the
crystal.  These errors show up in the $\ln(b^2)$ term in Equation
\ref{eqn:final_sol}, and probably add $\sim 8$\% uncertainty to
the measurements. Our best numbers for the absorption coefficients
are therefore $\alpha = (0.0034 \pm 0.0022) ~\mbox{m}^{-1}$  and
$\beta = (3.2\pm 0.5) \times 10^{-11}$ m/W.  These numbers do not
reflect the uncertainty in the thermal conductivity, which we do
not know. To our knowledge, this is the first determination of the
two-photon absorption coefficient at any wavelength in KNbO$_3$.

Earlier work on the linear absorption coefficient for KNbO$_3$
\cite{Busse94} found $\alpha = 0.001 \mbox{cm}^{-1} = 0.1
\mbox{m}^{-1}$ at 860 nm.  This value is much higher than this
work.  It may be that some of the absorption measured in previous
work was two-photon absorption.  That work used laser calorimetry
to measure the absorption coefficient at specific laser lines from
457 nm to 1064 nm.  At 860 nm, they focused a 900 mW laser beam
into a 5 mm crystal, and it is likely that the intensity was high
enough for multiphoton absorption to be important.

Our method for determining the one and two-photon absorption
coefficients is quite general.  It is independent of measurements
of light scattering inside the crystal.  It is also independent of
reflection measurements at the crystal faces.  This method is
particularly well suited to measurements of optical absorption
where the absorption coefficients are comparable to or smaller
than scattering and reflection coefficients.

\section{Transmission Lineshape With Blue Light}

When infrared light incident on the bowtie cavity is polarized
along the $b$-axis of the KNbO$_3$ crystal, it is possible to meet
the phase-matching conditions required for generating the second
harmonic at 423 nm.  The presence of blue light in the crystal
dramatically alters the transmission lineshape.  Not only is the
blue light itself absorbed by the crystal, it also significantly
increases the absorption of the infrared light.  This is called
``blue-light-induced-infrared-absorption'' (BLIIRA), and it has
been studied at length in the literature
\cite{Busse94,Goldberg93,Mabuchi94,Shiv95}.  Previous work has
found BLIIRA to be significant at blue light intensities down to
$7 \times 10^{-4}$ W/cm$^2$.  BLIIRA is minimized at longer
wavelengths and at higher crystal temperatures \cite{Shiv95}.
However, for our calcium work, we require high powers at 423 nm.

As before, we fix the laser frequency and scan the cavity length
through the resonance condition.  Both blue light at 423 nm and a
small amount of infrared light at 846 nm exits the cavity.  We
separate these wavelengths using dichroic mirrors, which transmit
95\% in the blue and reflect 99.5\% in the infrared.  To make sure
that no blue light reaches the infrared laser beam detector, we
use four of these mirrors in series in the infrared beam path
after the doubling cavity (see Figure \ref{fig:expLayout}).

In these experiments, we do not independently control the infrared
and blue beam intensities.  Rather, we optimize the crystal
temperature to maximize the blue light production for each
infrared power setting in the steady state.  Losses inside the
doubling cavity limit the maximum infrared power circulating
inside and therefore the maximum blue light produced.

\begin{figure}[t]
\centerline{\rotatebox{90}{\scalebox{.31}{\includegraphics{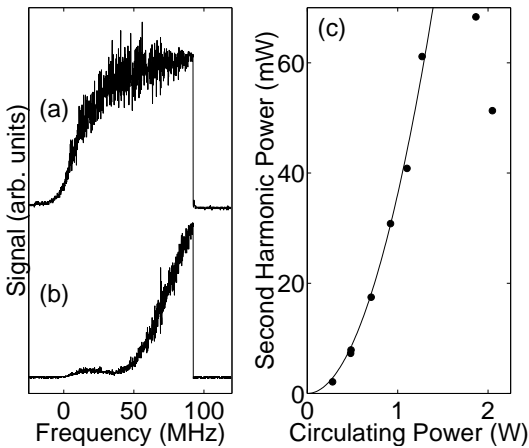}}}}
\caption{(a) Cavity transmission lineshape in the infrared at 846
nm when generating the second harmonic at 423 nm.  This signal is
taken from the highest intensities in the experiment. Note that
for low powers, the FWHM is 5 MHz.  (b) The second harmonic signal
for the same settings as \ref{fig:slrblue}a.  The shoulder to the
left of the signal is a Maker fringe (see text).  (c) Peak second
harmonic power as a function of peak infrared power.  For all but
the highest two power measurements, the peak blue power is given
by $P_{blue} = 36.2 \mbox{kW}^{-1}~P_{ir}^2$.
} \label{fig:slrblue}
\end{figure}

The transmission lineshape for the cavity when blue light is
present is shown in Figure \ref{fig:slrblue}a for our highest IR
powers. The temperature rise inside the crystal is significant. At
maximum power, as the cavity approaches the resonance condition,
only a small amount of blue light is initially generated because
the crystal temperature is too low to meet the phase matching
condition for second harmonic generation exactly (see Figure
\ref{fig:slrblue}b).  Closer to the cavity resonance condition,
the circulating power increases, the crystal heats up, and more
blue light is generated, which heats the crystal even more.  This
positive feedback continues, and the cavity demonstrates thermal
self-locking for up to 90 MHz, as shown in Figure
\ref{fig:slrblue}a.

The blue light signal in Figure \ref{fig:slrblue}b has a shoulder
on the low-frequency side of the maximum.  This is a Maker fringe.
As stated previously, we initially optimize the crystal
temperature for optimum blue light production in the steady state.
At the highest intensities, the temperature shift ($\sim 0.4
^{\mbox{\footnotesize o}}$C) due to thermal self-locking of the
cavity is a few times the temperature  phase matching bandwidth.

Because of these complicated thermal conditions, and especially
because the blue light production is not constant across the
transmission lineshape, we are reluctant to fit the lineshape to
any model for the highest laser intensities.  Instead, we find the
self-locking range graphically.  It is the falling step at the far
right hand side of the infrared transmission peak shown in Figure
\ref{fig:slrblue}.  This choice is valid as long as there is no
significant thermally-induced change in the cavity coupling
efficiency.  We can monitor these changes by measuring the light
reflected from the cavity.  For the relatively modest intensities
in this study, the minimum reflected light (which corresponds to
the maximum light inside the cavity) changes by only a few
percent, and only at the highest intensity measurements.  The
statistical scatter in the data is larger than this, and we
neglect this small systematic error.

\begin{figure}[t]
\centerline{\rotatebox{90}{\scalebox{.31}{\includegraphics{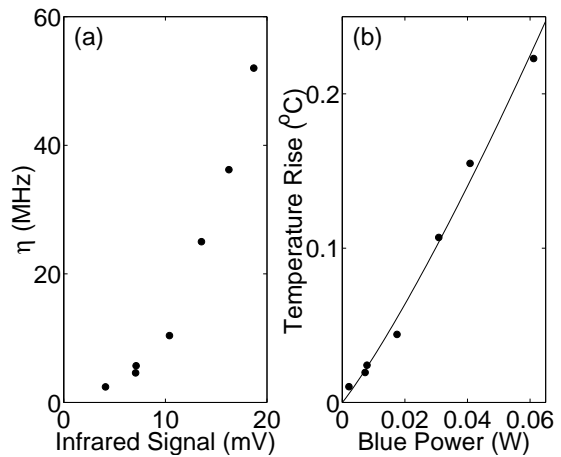}}}}
\caption{(a) Self-locking range, $\eta$, in the presence of blue
light versus infrared detector signal, excluding the highest power
data where the SHG process appears to be saturating.  (b) The data
from \ref{fig:horzpol}a with the temperature and power
calibrations, and with the temperature rise from the infrared
light subtracted off.  The solid line is a weighted fit to the
data (see text).
} \label{fig:horzpol}
\end{figure}

At lower laser intensities, where BLIIRA and other absorption
processes are less severe, we use a least-squares method similar
to Section 3 to find the self-locking range $\eta$.  The
self-locking range is plotted as a function of infrared laser
power in Figure \ref{fig:horzpol}a.  Note that except for the two
highest power measurements, the frequency doubling process is not
saturated (see Figure \ref{fig:slrblue}c).

With blue light present in the crystal, the temperature rise
inside the crystal has three sources: infrared absorption (one and
two photon), blue light absorption, and BLIIRA.  We can easily
subtract off the infrared contribution to the self-locking range
using the data in Figure \ref{fig:slr} as a look-up-table.  The
remaining contribution to the self-locking range is due to
temperature rise from blue light absorption and BLIIRA.

Because the blue light intensity and the temperature are not
constant along the length of the axis of the crystal, we must be
careful about how we define the temperature rise and the blue
light intensity.  We will define the absorption coefficient at 423
nm, $\delta$, by the equation

\begin{equation}
\Delta P = \int_{V} I_{b} \delta dV = \delta P_{b} L
\label{eqn:intensity}
\end{equation}

\noindent where $I_{b}$ is the blue light intensity inside the
crystal, $P_{b}$ is the measured blue power, and $L$ is the length
of the crystal.  In the non-depleted pump plane-wave
approximation, the blue light intensity grows quadratically with
distance $z$ in the crystal.  With the definition from Equation
\ref{eqn:intensity}, the blue light intensity can be written as
$I_{b}(r,z) = (12 P_{b} z^2 / \pi a^2 L^2) \exp(-4r^2/a^2)$, where
$a$ is the Gaussian waist for the infrared beam in the crystal and
we explicitly assume that the Gaussian waist for the blue beam is
$a/\sqrt{2}$.

The heat equation for blue light absorption and BLIIRA is written as

\begin{eqnarray}
\lefteqn{{1\over r}{\partial \over \partial r} \left( r {\partial
u\over
\partial r} \right) + {{ \partial^2 u}\over {\partial z^2}}} \nonumber \\
& = &
- z^2 \left( C e^{-4r^2/a^2} + D e^{-6 r^2/a^2} \right).
\label{eqn:heat2}
\end{eqnarray}

\noindent The first term on the right hand side of the equation is
the blue light absorption term where $ C = 12 P_{b} \delta / (\pi
a^2 L^2 K)$. The second term on the right hand side of the
equation is the BLIIRA term, where $D = 24 P_{b} P_{ir} \xi / (
\pi^2 a^4 L^2 K ) $, $P_{ir}$ is the measured infrared power,
assumed to be constant along the crystal length, and $\xi$ is the
BLIIRA absorption coefficient. As before, $u(r,z)$ is the
temperature inside the crystal and $K$ is the thermal
conductivity.

The solution of this equation for $r=0$ is

\begin{eqnarray}
\Delta T(z) & = & {{3 \delta }\over{ 4 \pi K }}
\left[\ln\left({{4b^2}\over{a^2}}\right) + \gamma \right]
{{z^2}\over{L^2}} P_{b} \nonumber \\
            & + & {{\xi}\over{a^2 \pi^2 K}}
\left[\ln\left({{6b^2}\over{a^2}}\right) + \gamma \right]
{{z^2}\over{L^2}} P_{b} P_{ir}, \label{eqn:ha}
\end{eqnarray}

\noindent but we have to be careful about how we define the
temperature rise. In Equation \ref{eqn:heat2}, the radial
derivatives are much larger than the longitudinal derivatives, and
the $\partial^2 u/\partial z^2$ term can be treated
perturbatively.  This approximation produces a temperature
function that increases quadratically with $z$ in the crystal.

We can write the temperature rise on axis as $\Delta T(z) = T^m
z^2/L^2$, where $T^m$ is the maximum temperature rise in the
crystal.  We can calculate the change in the optical pathlength
inside the crystal with this temperature profile.  It is equal to
the change in pathlength due to an {\sl equivalent} temperature
$\Delta T^{\prime} = T^m / 3$.  More expressly, we can write $T(z)
= T^m z^2 / L^2 = 3 T^{\prime} z^2 / L^2$, and rewrite Equation
\ref{eqn:ha} as

\begin{eqnarray}
T^{\prime} &=& {{\delta }\over{ 4 \pi K }}
\left[\ln\left({{4b^2}\over{a^2}}\right) + \gamma \right] P_{b}
\nonumber \\
&+& {{\xi}\over{3 a^2 \pi^2 K}}
\left[\ln\left({{6b^2}\over{a^2}}\right) + \gamma \right] P_{b}
P_{ir}, \label{eqn:haha}
\end{eqnarray}

Knowing this, we can convert the measured self-locking range of
the cavity in the presence of blue light into a temperature rise.
The data is plotted in Figure \ref{fig:horzpol}.  For this data,
the ``temperature rise'' is the average temperature rise in the
crystal, i.e.: $T^{\prime}$.  The ``blue power'' is the measured
blue power, i.e.: the power at the end of the crystal, $P_{b}$.

Again, we fit a function of the form

\begin{equation}
\Delta T = a_1 P_b + a_2 P_b P_{ir}
\end{equation}

\noindent to the data in Figure \ref{fig:horzpol}b using a
multivariable least-squares method and find $a_1 = 2.41
^{\mbox{\footnotesize o}}$C W$^{-1}$ and $a_2 = 1.04
^{\mbox{\footnotesize o}}$C W$^{-2}$.  Using equation
\ref{eqn:haha}, we determine the absorption coefficients $\delta =
13.3 ~ \mbox{m}^{-1}$ and $\xi = 2.2 \times 10^{-8} \mbox{m/W}$.
Apparently, there are no measurements in the literature of the
KNbO$_3$ absorption coefficient at 423 nm.  However, there are
measurements close to this wavelength \cite{Busse94}, and a value
of 13.3\% per cm is reasonable.  For this determination of the
absorption coefficient, we have the same kinds of uncertainties as
before, namely in the beam position in the crystal and in the
power measurement.  We estimate the error in this measurement to
be 15 \%.  Our final number for the absorption coefficient at 423
nm is $\delta = (13 \pm 2)$ m$^{-1}$.

This uncertainty estimate is perhaps conservative.  We have a
second KNbO$_3$ crystal.  In a simple transmission measurement at
423 nm, using the blue light after the frequency doubling cavity,
we find the linear absorption coefficient to be 13\% per cm, in
agreement with the work above.

We verified the validity of our perturbative approximation by
solving Equation \ref{eqn:heat2} numerically.  In the numerical
solution, the temperature profile is quadratic in $z$ except at
the very end of the crystal, where is flattens out slightly to
meet the boundary condition of no heat flowing out the crystal
face.  With our values of $\delta$, $\xi$, $a$, $P_{b}$, and
$P_{ir}$, the calculated temperature $T^{\prime}$ agrees with our
data to about one percent.

It is interesting that the above treatment of BLIIRA does not
distinguish between blue-light induced infrared absorption and
infrared light induced blue-light absorption.  This model of
BLIIRA only postulates that there is a two-color two-photon
absorption cross-section. This treatment of two-color two-photon
absorption in KNbO$_3$ is perhaps simplistic.  However, it serves
the purpose in this paper of demonstrating that thermal
self-locking of a Fabry-Perot cavity can be used to sensitively
characterize the linear and nonlinear optical properties of an
absorbing crystal.

\section{New Measurements of ${{dn}/{dT}}$}

From the calibration measurements that convert $\eta$ in Figures
\ref{fig:slr} and \ref{fig:horzpol} to a temperature rise, we can
extract the temperature-dependent change in the index of
refraction for KNbO$_3$.  When the crystal temperature changes,
the optical pathlength of the bowtie cavity also changes because
both the crystal length and the index of refraction depend on
temperature:

\begin{equation}
{{{d\cal{L}}}\over{dT}} = L_{c} \left[{dn_{x}\over dT}+\alpha_T \left(
n_{x}-n_{0}\right) \right],
\label{eqn:opl}
\end{equation}

\noindent where ${\cal L}$ is the cavity optical pathlength, $L_c$
is the crystal length (10 mm), $\alpha_T$ is the coefficient of
linear expansion for KNbO$_3$, and $n_x$ and $n_0$ are the indices
of refraction for the crystal and air, respectively.

In a relatively simple measurement, we can determine $dn/dT$ for
the crystal across a wide range of temperatures. We fix the laser
frequency and the cavity mirror positions.  We attenuate the laser
intensity to 10$\mu$W to avoid thermal self-locking.  The crystal
temperature can be independently controlled because it is mounted
on a Peltier device.  We measure the cavity transmission while
slowly sweeping the crystal temperature over several degrees.  At
certain temperatures, the cavity optical pathlength meets the
resonance condition, and we measure a peak in the cavity
transmission.  The temperature between transmission peaks in these
measurements corresponds to changing the optical pathlength by one
optical wavelength, 846 nm.  We repeat these measurements over a
wide temperature range to determine $d{\cal L}/dT$ from -12
$^{\mbox{\footnotesize o}}$C to about 50 $^{\mbox{\footnotesize
o}}$C.

\begin{figure}[t]
\centerline{\rotatebox{90}{\scalebox{.32}{\includegraphics{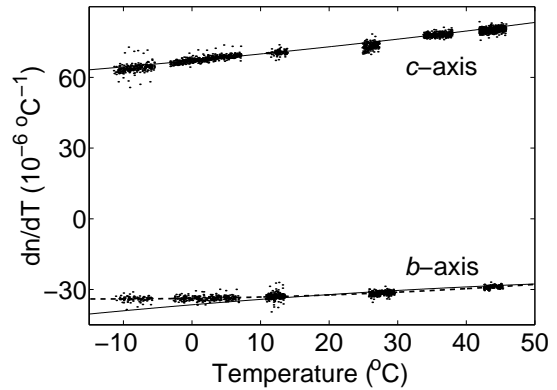}}}}
\caption{The change in the index of refraction with temperature
for light polarized along the $b$ and $c$-axes.  The points are
from this work.  The solid line is from Ghosh \cite{Ghosh94}.  The
dashed line is a second order polynomial fit to the data.
} \label{fig:ghosh}
\end{figure}

Using equation \ref{eqn:opl}, we convert $d{\cal L}/dT$ to
$dn/dT$.  With $n_c = 2.13$, $n_b = 2.28$, $\alpha_T = 5 \times
10^{-6}~ ^{\mbox{\footnotesize o}} \mbox{C}^{-1}$, and $d{\cal
L}/dT = \lambda/ \Delta T$, we determine $dn/dT$ for several
temperatures. Our values are shown in Figure \ref{fig:ghosh} for
light polarized along the $c$-axis and $b$-axis, as labeled. Also
shown in the figure is the data from Ghosh \cite{Ghosh94}. The
line from Ghosh is a fit to his measurements of the index of
refraction between 0 and 140 $^{\mbox{\footnotesize o}}$C.  We
extrapolate that fit to lower temperatures to compare with our
data.

For light polarized along the $c$-axis, our measurements agree
with those of Ghosh over our entire temperature range.  For light
polarized along the $b$-axis, in the range 5 to 50
$^{\mbox{\footnotesize o}}$C, which is roughly the range for which
our data overlaps Ghosh's data, our measurements also agree.
However, below $\sim 5^{\mbox{\footnotesize o}}$C, our data
departs slightly from Ghosh's.

\begin{figure}[b]
\rotatebox{90}{\scalebox{.32}{\includegraphics{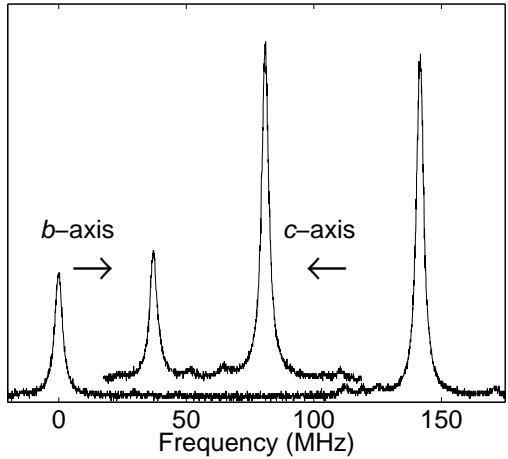}}}
\caption{Transmitted lineshape  as a function of cavity length for
two temperatures. The upper trace has been displaced for clarity.
The polarization of the light incident on the cavity is rotated a
few degrees relative to the $c$-axis.  The larger feature is
polarized along the $c$-axis, and the smaller feature is polarized
along the $b$-axis, as labeled.  This shows the magnitude and
direction of the change in the cavity resonance length with
temperature.
} \label{fig:reldndt}
\end{figure}

To check our data, we made an additional measurement.  When the
polarization of the light incident on the cavity is rotated
slightly relative to the $c$-axis, the transmission lineshape has
two features, one for each polarization.  Because the index of
refraction is different for the two polarizations ($b$ and $c$),
the optical pathlength is also different, and each polarization is
transmitted at a different cavity length.  When we change the
temperature of the crystal, the two peaks in the transmission
lineshape shift by different amounts in different directions, as
shown in Figure \ref{fig:reldndt}.  We measured these shifts for a
set of small temperature changes near -11.5 $^{\mbox{\footnotesize
o}}$C and determined the relative changes in the indices of
refraction, $\left(dn_c/dT\right) / \left(dn_b/dT\right)$.  These
measurements confirm our low temperature data in Figure
\ref{fig:ghosh}. Following the treatment of Ghosh, we fit our
$dn/dT$ data for the $b$-axis to a second-order polynomial, and
find

\begin{eqnarray}
{{dn_b}\over{dT}} &=&  1.55809 \times 10^{-9} T^2 \nonumber \\
                  &+&  4.06912 \times 10^{-8} T
- 33.76305 \times 10^{-6}
\end{eqnarray}

\noindent where the units of $dn/dT$ are $^{\mbox{\footnotesize
o}}$C$^{-1}$, and $T$ is measured in Celsius.

\section{Conclusion}

We have demonstrated how thermal self-locking of a Fabry-Perot
cavity can be used to determine optical properties of nonlinear
crystals.  We have determined new values for the one- and
two-photon absorption coefficients in KNbO$_3$ at 846 nm.  We have
also determined new values for the one photon (linear) absorption
coefficient at 423 nm, and a BLIIRA absorption coefficient under
specific circumstances.  Finally, we present new measurements of
the temperature-dependent change in the indices of refraction for
the $b$ and $c$ axes.

These measurements all derive from the thermal response of an
absorbing Fabry-Perot cavity to light circulating in the cavity.
The methods are general and can be widely applied to other
systems, particularly those in which the absorption coefficients
are small compared to scattering and reflection coefficients.

\section{Acknowledgements}

We express our appreciation to Ross Spencer for his computational
assistance.   This work is supported in part by a grant from the
Research Corporation and from the National Science Foundation
under Grant No. PHY-9985027.

\end{document}